\documentclass[
aps,
pra,
superscriptaddress,
preprintnumbers,
10pt
]{revtex4-1}

\usepackage[utf8]{inputenc}  
\usepackage[T1]{fontenc}
\usepackage[english]{babel}  

\usepackage{amsmath,amssymb}
\usepackage{hyperref}
\usepackage{verbatim}
\usepackage{graphicx}

\usepackage{tabularx}
\usepackage{booktabs}
\usepackage{multirow}
\usepackage{bm}
\usepackage{bbold}
\usepackage{mathtools}

\usepackage{color}
\definecolor{dred}{rgb}{.8,0.2,.2}
\definecolor{ddred}{rgb}{.8,0.5,.5}
\definecolor{dblue}{rgb}{.2,0.2,.8}

\newcommand{\Ciso}{$^{13}$C\ }
\newcommand{\Cis}{$^{12}$C\ }

\newcommand{\Nis}{$^{15}$N\ }
\newcommand{\SWAP}{\textsc{swap}}
\newcommand{\DYNAMO}{\textsc{dynamo}}

\DeclareMathOperator{\trace}{Tr}

\DeclareMathOperator{\Ker}{Ker} 
\DeclareMathOperator{\Supp}{Supp} 

\DeclareMathOperator{\diag}{diag}

\newcommand{\be}{\begin{equation}}
\newcommand{\ee}{\end{equation}}

\newcommand{\bal}{\begin{align}}
\newcommand{\eal}{\end{align}}
\newcommand{\bea}{\begin{eqnarray}}
\newcommand{\eea}{\end{eqnarray}}
\newcommand{\bpm}{\begin{pmatrix}}
\newcommand{\epm}{\end{pmatrix}}

\newcommand{\I}{\mathbb{1}} 

\newcommand{\ket}[1]{\ensuremath{\left| #1 \right \rangle}}
\newcommand{\bra}[1]{\ensuremath{\left \langle #1 \right |}}

\newcommand{\ketbra}[2]{\ket{#1}\bra{#2}}

\newcommand{\gate}[1]{{\sc #1}}

\begin{document}

\title{High fidelity spin entanglement using optimal control}

\author{Florian Dolde}
\thanks{These authors contributed equally.}
\affiliation{3rd Institute of Physics, Research Center Scope and IQST, University of Stuttgart, 70569 Stuttgart, Germany}

\author{Ville Bergholm}
\thanks{These authors contributed equally.}
\affiliation{Institute for Scientific Interchange, Via Alassio 11/c, 10126 Torino, Italy}

\author{Ya Wang}
\thanks{These authors contributed equally.}
\affiliation{3rd Institute of Physics, Research Center Scope and IQST, University of Stuttgart, 70569 Stuttgart, Germany}

\author{Ingmar Jakobi}
\affiliation{3rd Institute of Physics, Research Center Scope and IQST, University of Stuttgart, 70569 Stuttgart, Germany}


\author{Sebastien Pezzagna}
\affiliation{Institute for Physics, University of Leipzig, 04103  
Leipzig, Germany}

\author{Jan Meijer}
\affiliation{Institute for Physics, University of Leipzig, 04103  
Leipzig, Germany}


\author{Philipp Neumann}
\affiliation{3rd Institute of Physics, Research Center Scope and IQST, University of Stuttgart, 70569 Stuttgart, Germany}

\author{Thomas Schulte-Herbr\"uggen}
\affiliation{Department of Chemistry, Technical University Munich, 85747 Garching, Germany} 

\author{Jacob Biamonte}
\affiliation{Institute for Scientific Interchange, Via Alassio 11/c, 10126 Torino, Italy}

\author{J\"org Wrachtrup}
\email{j.wrachtrup@physik.uni-stuttgart.de}
\affiliation{3rd Institute of Physics, Research Center Scope and IQST, University of Stuttgart, 70569 Stuttgart, Germany}

\date{\today}

\begin{abstract}
Precise control of quantum systems is of fundamental importance for
quantum device engineering, such as is needed
in the fields of quantum information
processing, high-resolution spectroscopy and quantum metrology.
When scaling up the quantum registers in such devices, several challenges arise:
individual addressing of qubits in a dense spectrum while
suppressing crosstalk, creation of entanglement between distant nodes,
and decoupling from unwanted interactions.
The experimental implementation of optimal control is a prerequisite to meeting these challenges.
Using engineered microwave pulses, we experimentally demonstrate
optimal control of a prototype solid state spin qubit system
comprising thirty six energy levels.
The spin qubits are associated with proximal nitrogen-vacancy~(NV) centers in diamond.
We demonstrate precise single-electron spin qubit operations with an
unprecedented fidelity $F\approx 0.99$ in combination with
high-efficiency storage of electron spin states in a nuclear spin
quantum memory.
Matching single-electron spin operations with spin-echo techniques, we
further realize high-quality entangled states (F > 0.82) between two
electron spins on demand.
After exploiting optimal control, the fidelity is mostly limited by
the coherence time and imperfect initialization. Errors from
crosstalk in a crowded spectrum of 8 lines as well as detrimental
effects from active dipolar couplings have been simultaneously
eliminated to unprecedented extent.
Finally, by entanglement swapping to nuclear spins, nuclear spin entanglement over a length scale of $25$~nm is demonstrated.
This experiment underlines the importance of optimal control for
scalable room temperature spin-based quantum information devices.
\end{abstract}

\maketitle

\parindent 0mm
\parskip 5mm


High fidelity quantum operations, including gates, on demand entangled
state generation and coherent control in general,
represent a fundamental prerequisite for all quantum information technologies
such as error correction, quantum metrology and of course quantum information processing, wherein the
hardware and its control must satisfy the DiVincenzo criteria~\cite{divincenzo_physical_2000}.
A very promising class of quantum information devices are spin qubits
in solids, like phosphorus in silicon~\cite{pla_high-fidelity_2013},
rare earth ions in a solid state matrix~\cite{yin_optical_2013},
quantum dots~\cite{le_gall_optical_2011}
and defects in diamond or silicon carbide~\cite{dolde_room_temperature_2013, weber_defects_2011}.
Although there have been recent experimental advances in increasing the
number of coherently interacting qubits implemented using these
technologies, gate quality has been
limited~\cite{Pezzagna_mica_2011,dolde_room_temperature_2013}.
Optimal control,
often seen as a central tool for turning principles of quantum theory
into new technology~\cite{Vandersypen_NMR_optimal_2005},
seems to be the only practical way to
ensure functionality even in light of device imperfections, and to
overcome several impactful features found when scaling up the register size
such as unwanted crosstalk between control fields designed for
individual qubit control.
It is gradually being exploited in many other experimental settings,
including ion traps~\cite{BW08}, optical lattices~\cite{BDZ08},
solid-state devices~\cite{Nak03, CCG09, CW_Nat08}, and NMR~\cite{control-NMR-encyc}.

\begin{figure}
  \includegraphics[width=1.0\textwidth]{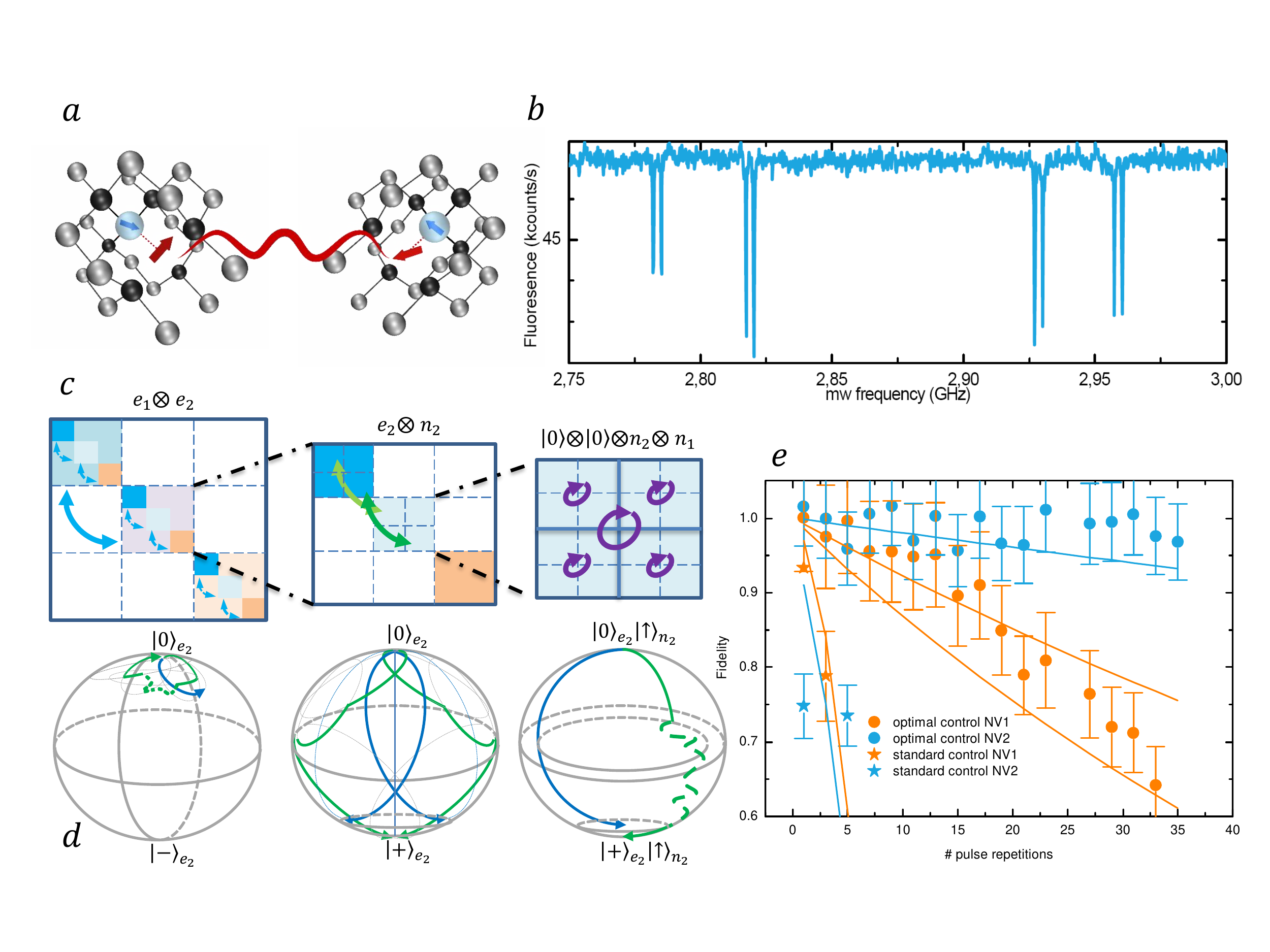}
  \caption{\label{fig:1}
    \textbf{Optimal control of a single qutrit.}
		(a) Schematic of the NV-NV pair used in this work.
    (b) Optically detected magnetic resonance (ODMR) spectrum of the NV pair.
		The outer pairs of transitions correspond to NV1 and the inner pairs to NV2.
		The splitting within one pair of $\approx 3\,$MHz is due to the hyperfine coupling with the \Nis nucleus.
		Spin transitions of separate NV centers are separated by $\approx30\,$MHz.
    (c) Concatenated exemplary representation of the 36 dimensions of the Hilbert space $e_1 \otimes e_2 \otimes n_2 \otimes n_1$ corresponding to the two coupled NV centers.
		The left panel, subsystem $e_1 \otimes e_2$ is shown with blue arrows illustrating electron spin manipulation on NV1 (bold, solid arrows) and its crosstalk on NV2 (dashed arrows).
		The center panel shows the subspace $\ket{0}_{e_1}\otimes e_2 \otimes n_2$ and green arrows of different tones illustrate the detuning due to hyperfine interaction.
		The right panel, the always-on nuclear spin precession by external static magnetic fields in subspace $\ket{0}_{e_1}\otimes \ket{0}_{e_2} \otimes n_2 \otimes n_1$ is illustrated (curved arrows).
    (d) Schematic Bloch sphere representation of the action of standard control (blue) and optimal control (green) considering the above mentioned effects.
		(left) Manipulation of spin $e_1$ should not affect the  state of spin $e_2$ via crosstalk,
		(center) despite hyperfine interaction the spin $e_2$ should be inverted regardless of the state of nuclear spin $n_2$, and
		(right) always on rotation of nuclear spins $n_1,n_2$ for electron spin states $\ket{0}_{e_1},\ket{0}_{e_2}$ should be avoided if not exploited.
(e) Repeated application of a \gate{not} gate targeted on spin $e_1$, implemented using a standard $\pi$-pulse (stars) as compared to an optimized gate (filled circles).
With an odd number of applications the effect
should always be the same (spin flipped for $e_1$, unchanged for $e_2$).
The fidelity with respect to these target states is displayed for both spins (orange and blue).
Where optimal control pulses allow for at least 20 repetitions without
a significant loss of fidelity and negligible crosstalk within our
measurement error, $\pi$-pulses show low fidelity and strong crosstalk
already after the first gate application.
  }
\end{figure}

Here we demonstrate a decisive step towards overcoming the aforementioned challenges by developing
optimal control methods for solid state spin registers to dramatically increase their utility.
We explore a prototype of such a quantum register, operating at ambient conditions, based on two neighboring NV centers in diamond.
Our register is built of two electron spin qutrits plus two nuclear spin qubits, realizing thirty six energy levels in total (see Fig.~\ref{fig:1}a).
We implemented optimal control on these thirty six levels to realize a
fully functional four qubit register with high fidelity entanglement
between electron and nuclear spins.

NV centers in diamond are unique and interesting solid state devices for implementing quantum technologies~\cite{kubo_strong_2010,steinert_magnetic_2013,staudacher_nuclear_2013,mamin_nanoscale_2013,grinolds_nanoscale_2013,kucsko_nanometer_2013,neumann_high-precision_2013-1,toyli_fluorescence_2013-2,dolde_electric-field_2011,dutt_quantum_2007,togan_quantum_2010,bernien_heralded_2013}.
More precisely, the NV center's electron spin, the nitrogen nuclear
spin and proximal \Ciso nuclear spins form a small quantum register
which is the fundamental building block of potential NV-based quantum
devices.
Even without optimal control, several hallmark demonstrations of their properties have been possible, including coherent single qubit operation and readout \cite{gruber_scanning_1997,jelezko_observation_2004-1}, controlled qubit gates \cite{jelezko_observation_2004,dutt_quantum_2007} and entanglement generation \cite{neumann_multipartite_2008,dolde_room_temperature_2013,togan_quantum_2010,bernien_heralded_2013} at ambient conditions.
NV centers have been shown to exhibit coherence times on the order of milliseconds in isotopically purified diamond~\cite{balasubramanian_ultralong_2009} or through dynamical decoupling~\cite{naydenov_dynamical_2011,lange_universal_2010}.
This has to be compared to coherent control in the nanosecond regime~\cite{fuchs_gigahertz_2009}.
Particularly, the nuclear spins have proven to be a valuable resource for high fidelity readout \cite{neumann_single-shot_2010,dreau_single-shot_2013} and as a non-volatile memory \cite{maurer_room-temperature_2012}.

The effective magnetic dipolar interaction range of spins is limited by the coherence life time (here: $\sim 2\,$ms) to about
$\sim 50$~nm for electron spins and $\sim 5$~nm for nuclear ones.
In contrast to optical techniques,
microwaves cannot be focused down to these length scales.
Hence the addressability needs to be achieved by
separating the spins' resonance frequencies sufficiently, for example by spatially modulating the local, static magnetic field (see Fig.~\ref{fig:1}b).
Electron and nuclear spins are then individually and coherently manipulated by mw and rf fields.
However, the dense spectrum leads to non-negligible crosstalk (see Fig.~\ref{fig:1}c).
Nuclear spins can be additionally controlled via their hyperfine interaction with the neighboring electron spin.
Effectively the electron spin state sets the axis and speed of nuclear Larmor precession \cite{hodges_universal_2008}.
Please note that the dipolar interactions among spins are always on.
Consequently, it becomes challenging to separate single-qubit and controlled qubit gates (see Fig.~\ref{fig:1}c)
While this is a minor issue for standard spectroscopy techniques, the
fidelity of the gates can be drastically affected, especially for
repeated gate application.

Improving gate fidelity is a nontrivial task; the main reason for this
being the high spectral density.
The interaction of an applied microwave field with a spin can be described by the Rabi formula
\begin{equation}
p_{\mathrm{target}}(t) = \frac{\Omega^2}{\Omega^2+\Delta^2}
\sin^2{\frac{\sqrt{\Omega^2+\Delta^2}\, t}{2}}\, ,
\label{eq:Rabi_formula}
\end{equation}
giving the probability $p_{\mathrm{target}}$ for a spin flip into a target state.
Here the Rabi frequency~$\Omega$ is the strength of the applied mw
field and $\Delta$~is the detuning of the mw frequency from the actual
spin transition.
Apparently, high fidelity control of a single transition (i.e. $p_{\mathrm{target}}\approx 1$) can be achieved by a large ratio $\Omega/\Delta$.
However, in the case of single-qubit gates on the electron
spin (i.e. irrespective of the nuclear spin state), the hyperfine
interaction sets a lower bound for the detuning~$\Delta$ and the
spectral density sets an upper bound for Rabi frequency~$\Omega$ in
order to avoid crosstalk.
In our particular case the hyperfine interaction is $\approx 3\,$MHz and the spectral separation of individual NV transitions is $\approx 30\,$MHz (see Fig.~\ref{fig:1}b).
This limits the fidelity of a single-qubit \gate{not} gate to $F \approx 0.9$  (see Fig.~\ref{fig:1}c).
During the finite duration of electron spin control, additionally, the nuclear spins undergo rotations dependent on the respective electron spin projection.
While this will be exploited for nuclear spin control (see below) it further reduces the fidelity of pure electron spin gates  (see Fig.~\ref{fig:1}c).

For designing high fidelity experiments, optimal control methods are gradually
establishing themselves as valuable means to get the most out of an actual
quantum experimental setting~\cite{DowMil03,WisMil09}.
The general scenario involves minimizing a cost functional
under the constraint that the system follows a given equation of
motion. For state transfer or quantum gate synthesis in a closed
system (neglecting decoherence for the moment)
this amounts to the controlled Schrödinger equation.
The control sequence is usually taken to be piecewise constant
so the pulse shapes can easily be fed to a digital pulse shaper.
A convenient error function for quantum gates is the infidelity
$E(U) = 1-\frac{1}{D} \left|\trace(U_{\text{target}}^\dagger U)\right| \in [0, 1]$,
which absorbs unphysical global phases.
With these stipulations,
effecting a desired quantum gate is (in principle) a standard task
that can be conveniently addressed e.g.~by the GRAPE~\cite{GRAPE} optimization
algorithm in the \DYNAMO{} numerical
optimal control toolbox~\cite{dynamo}.
To handle non-idealities like crosstalk we use a modified rotating wave approximation (RWA),
taking sufficiently slowly rotating Hamiltonian components into
account in addition to the static ones (see Supplementary Information).
Our optimization framework also allows for Markovian~\cite{JPB_decoh} and
non-Markovian~\cite{PRL_decoh2} relaxation to be included.


Our experimental system consists of two $^{15}$NV centers
separated by a distance of $25 \pm 2$~nm,
with an effective mutual dipolar coupling of
$\nu_{\text{dip}} = 4.93 \pm 0.05$~kHz \cite{dolde_room_temperature_2013} (see Fig.~\ref{fig:1}a).
Each NV center has an electron spin-1
(denoted $\vec{S}$) and a \Nis nuclear spin-1/2 (denoted $\vec{I}$),
hence the system  exhibits $(3 \cdot 2)^2 = 36$ energy levels in total.
We label the eigenstates $m_S=+1,0,-1$ of the $S_z$ spin operator 
with the symbols~$(+,0,-)$, and the eigenstates $m_I=+1/2,-1/2$ of~$I_z$
with~$(\uparrow, \downarrow)$.
Although the electron spin is a qutrit we use the states $\ket{\pm}$
as an effective qubit and $\ket{0}$ as an auxiliary state.
Since individual optical
addressing is challenging at this short a distance, the
readout is performed simultaneously on both NV centers.
Individual addressing of both NVs' spin transitions is realized by different crystal field directions and proper magnetic field alignment
resulting in a spectral separation of $\approx 30$~MHz between the individual
NV transitions (see Fig.~\ref{fig:1}b).
Despite the misaligned magnetic field, spin states $\ket{\pm},\ket{0}$ remain approximate eigenstates because of the much stronger crystal field along the NV axis.
The hyperfine interaction of spin states $\ket{\pm}$ with the
\Nis nuclear spin aligns the latter along the NV axis  and splits $\ket{\uparrow}$ and $\ket{\downarrow}$ by $3.01$~MHz allowing for electron spin operations controlled
by the nuclear spin (see Fig.~\ref{fig:1}b).
While the product states $\ket{ \left\{+,-\right\} }\otimes\ket{ \left\{\uparrow,\downarrow\right\}}$ are eigenstates and form the computational basis of each individual NV center, the auxiliary states $\ket{0}\otimes \ket{ \left\{\uparrow,\downarrow\right\}}$ are not eigenstates and therefore facilitate electron spin controlled nuclear spin rotations.
Please note that the spin transition frequencies of the two individual NV centers are sufficiently far detuned ($30\,$MHz) to avoid mutual flip-flop dynamics induced by the dipolar interaction ($5\,$kHz).
Instead, a decoupling sequence is used to realize a controlled phase gate among the two NV centers (see Fig.~\ref{fig:3}a).
The dephasing times of NV1 are
$T_{\text{2dq}}^* = 27.8 \pm 0.6~\mu$s
and $T_{\text{2dq}} = 150 \pm 17~\mu$s, and
those of NV2 are
$T_{\text{2dq}}^* = 22.6 \pm 2.3~\mu$s and
$T_{\text{2dq}} = 514 \pm 50~\mu$s.


Prior to implementing optimal control, a proper characterization of the spin Hamiltonian and the control field is necessary.
In particular the response of the NV electron spin to different frequencies and amplitudes of the control field is calibrated, compensating for nonlinearities and spectral inhomogeneities.
To compare standard and optimal control, we repeatedly apply a \gate{not} gate to the electron spin of NV1 interrupted by a small free evolution time $\left(\left[\pi_{\text{optimal/standard}}-\tau_{\text{free evolution}} - \right]^{2k+1}\right)$ (see Fig. \ref{fig:1}e).
First, the system is initialized into state $\ket{m_S^{\text{NV1}},m_S^{\text{NV2}}}=\ket{00}$.
If the applied gate is perfect, the state of NV1 always results in $\ket{+}$ and that of NV2 in $\ket{0}$, neglecting decoherence.
However for standard control with rectangular time-domain pulses with $\Omega_{\text{Rabi}} = 10\,$MHz, the experimental results
show a fast decay of population in $\ket{+}$ for NV1 and a strong crosstalk effect on NV2 (i.e. decrease of population in $\ket{0}$) (Fig.~\ref{fig:1}e).
In contrast, for optimal control the decay is much slower and almost no crosstalk is observed for 35 applications of the \gate{not} gate.
To quantify the precision of optimal control, we use a randomized benchmarking protocol and assume independent error sources for all applied optimal gates.
A fidelity between $0.9851$ and $0.9920$ for the optimal \gate{not} gate on NV1 and $0.9985$ for the identity gate on NV2 are achieved by fitting the experimental results.

\begin{figure}
  \includegraphics[width=0.5\textwidth]{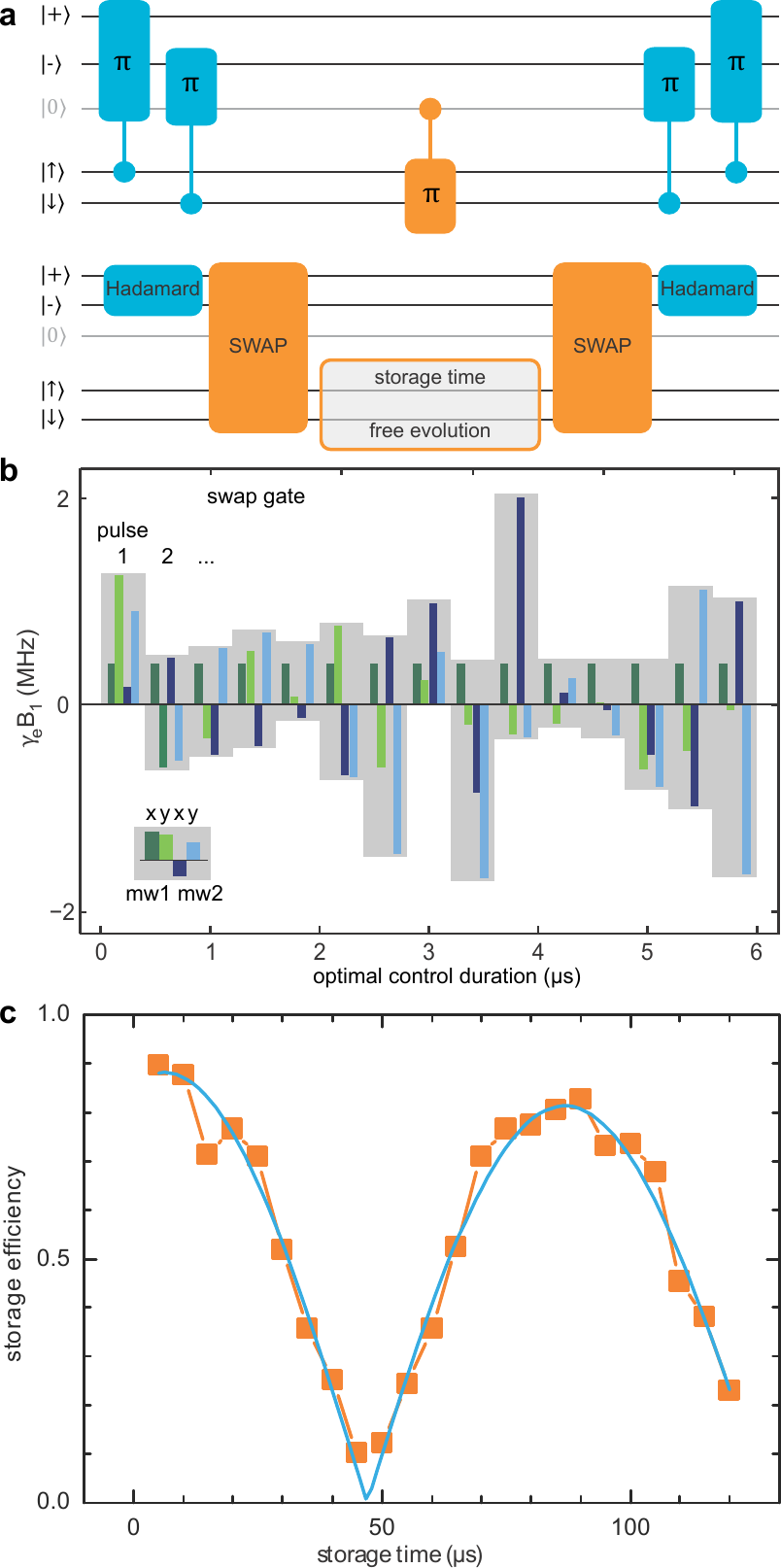}
  \caption{\label{fig:2}
    \textbf{\SWAP{}~gate between electron spin and nuclear spin.}
(a) Quantum wire diagrams for (top) \SWAP{} gate between qubits
    $\ket{\pm}$ and $\ket{\uparrow \downarrow}$ via standard control,
    utilizing the auxiliary state $\ket{0}$, and (bottom)
    creation, storage, retrieval and readout of a superposition state employing an optimized \SWAP{} gate.
(b) Optimal control \SWAP{} gate consisting of fifteen rectangular pulses (gray bars) each $0.4\,\mu$s long.
    Each pulse has two frequency components, corresponding
    to transitions $\ket{0} \leftrightarrow \ket{+}$ (mw1, green) and $\ket{0} \leftrightarrow \ket{-}$ (mw2, blue).
    In addition, each frequency component (mw1, mw2) has an in-phase and an out-of-phase amplitude (dark, bright).
    All four contributions to a single pulse are applied simultaneously during the whole pulse duration.
(c) The retrieved superposition state reveals the free evolution during quantum state storage.
		Here we show the $\left| \left \langle \hat{I}_x \right \rangle \right|$ component of the stored coherence.
		Apparently, the Larmor precession of the nuclear spin superposition state leads to a phase accumulation.
  }
\end{figure}
\begin{figure}
  \includegraphics[width=0.5\textwidth]{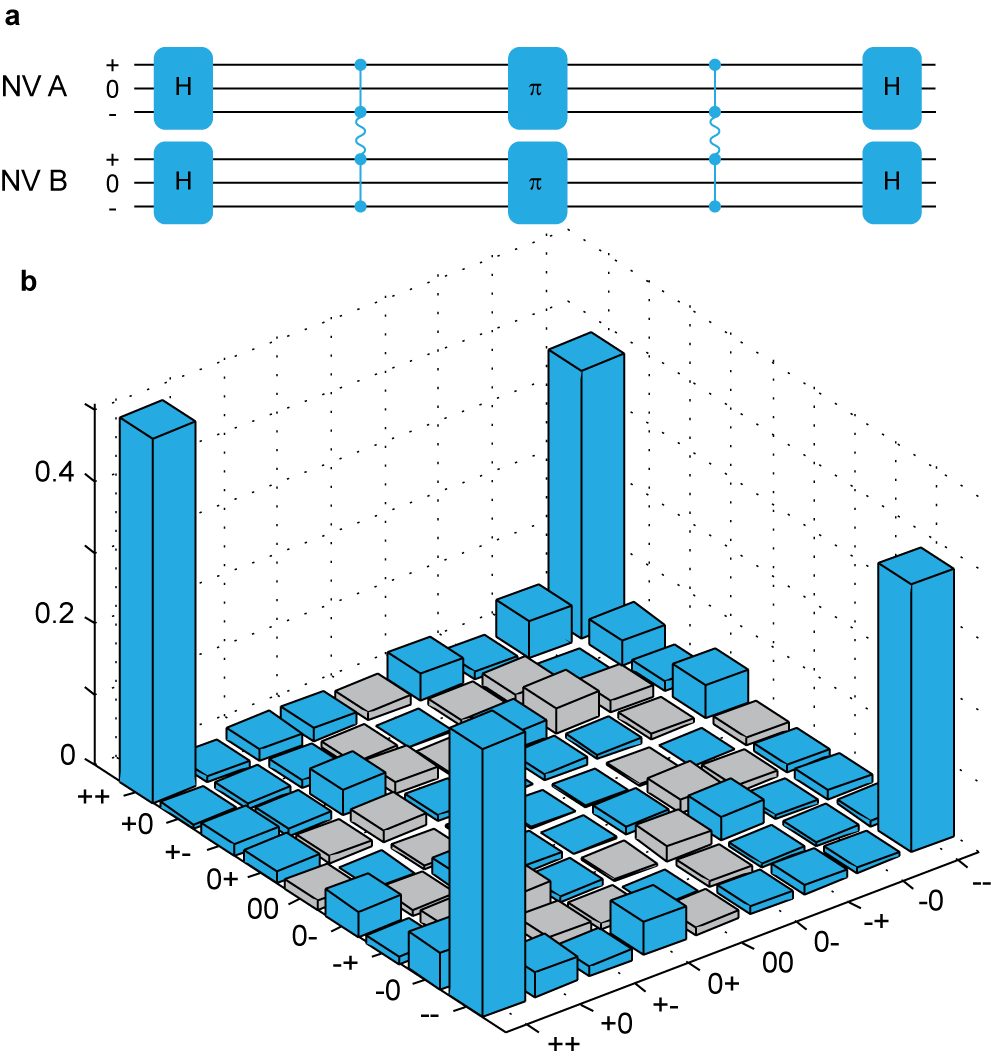}
  \caption{\label{fig:3}
    \textbf{Electron spin entanglement.}
    (a) Quantum wire diagram of the entanglement sequence.
    (b) Density matrix of the created Bell state
    $\ket{\Phi_{\text{dq}}} = \ket{++}+i\ket{- -}$ (F=$0.824\pm0.015$).
		The blue columns represent measured values.
		Please note that except for the main four columns representing the entangled state and the entries $\ketbra{+-}{+-}$ and $\ketbra{-+}{-+}$ all values are consistent with shot noise of the measurement process.
		The gray columns are upper bounds given by the measured main diagonal entries and the requirements for a physical state.
  }
\end{figure}

The \Nis nuclear spins couple to magnetic fields much more weakly than the NV electron spins, and consequently have much longer coherence times.
Therefore they are ideal long lived storage qubits
\cite{Maurer_Science}, which are easily integrated into a qubit
register via their hyperfine coupling to the electron spin.
Various methods have been worked out for controlled nuclear spin operations.
A particularly convenient one utilizes hyperfine interaction between electron and nuclear spins.
To this end the state $\ket{0}$ acts as an ancilla level for nuclear spin manipulation.
In contrast to $\ket{\pm}$, state $\ket{0}$ exhibits no hyperfine coupling to the nuclear spin.
Therefore, in state $\ket{0}$ the nuclear spin is mainly susceptible
to the external magnetic field and consequently undergoes Larmor
precession around it with the angular frequency
$\omega_L=\gamma_N \sqrt{B_{0_{\parallel}}^2 + \eta B_{0_{\perp}}^2}$,
where $\gamma_N$ is the nuclear gyromagnetic ratio.
More precisely, the later field is an effective one, where $\eta$ describes the enhancement due to dressed nuclear-electron spin states \cite{childress_coherent_2006, hodges_universal_2008}.
In the current experiment this effective field is almost perpendicular to the NV axis (see methods).
Therefore the precession is a coherent oscillation between states $\ket{\uparrow}$ and $\ket{\downarrow}$ which realizes a fast \gate{crot} (controlled rotation) gate  on the nuclear spin. 
Having at hand \gate{crot} gates for electron and nuclear spins we can design a \SWAP{} gate for quantum information storage.
The standard approach is a complex pulse sequence (Fig.~\ref{fig:2}a).
However, the imperfections of each operation will accumulate and largely reduce the performance of the gate.
We define the storage efficiency as the ratio of qubit coherences after and before storage and retrieval. 
For the standard \SWAP{} gate we found the storage efficiency to be Eff$_{\text{std}}=0.50 \pm 0.07$, which is mainly limited by crosstalk.
With optimal control we tailored a \SWAP{} gate
(exchanging the states $\ket{+\uparrow}$ and~$\ket{-\downarrow}$)
with a significantly better performance compared to the standard approach (Fig.~\ref{fig:2}c).
A storage efficiency of Eff$_{\text{opt}}=0.89 \pm 0.01$ was measured. Eff$_{\text{opt}}$ is limited by decoherence during the \SWAP{} operation.
The oscillation of the storage efficiency shown in figure \ref{fig:2}c is due to free evolution of the nuclear spin coherence $e^{-i \omega_nt}$ during the storage period, which can be accounted for.


So far we have demonstrated coherent control within one NV center node.
However, scalability arises from coherent interaction of neighboring NV nodes.
The two NV centers of our register interact very weakly compared to their mutual detuning owing to Zeeman interaction.
Thus they only influence the phase accumulation on the other NV.
To generate an entangled state we therefore design and apply a controlled phase gate. 
Specifically, after initialization to \ket{00}, a local superposition state
$\ket{+} +\ket{-}$ is created on both NV centers.
Free evolution under the
$H_{\text{int}}/\hbar = 2 \pi \nu_{\text{dip}} S_z \otimes S_z$ term of the Hamiltonian
will then make the states accumulate a relative phase
$\phi := 4 \pi \nu_{\text{dip}} \tau$, where
$\tau$ is the evolution time,
effecting a non-local phase gate which entangles the electron spin states.
$\tau = \frac{1}{8 \nu_{\text{dip}}} \approx 25.4$~$\mu$s
will yield
$\phi = \pi/2$, at which point the state can be
locally mapped into the Bell-type entangled state~$\ket{\Phi_{\text{dq}}}$:
\bea
\ket{00} & \xrightarrow{U_1 \otimes U_1} & \frac{1}{2}\left(\ket{+}+\ket{-}\right) \otimes \left(\ket{+}+\ket{-}\right) \nonumber \\
& \xrightarrow{e^{-i H_{\text{int}} \tau/\hbar}} & \frac{1}{2} \left((\ket{++}+\ket{--}) +e^{i \phi}(\ket{+-}+\ket{-+})\right)\\
& \xrightarrow{U_3 \otimes U_3} & \frac{1}{\sqrt{2}}\left(\ket{++}+i\ket{--}\right)
=: \ket{\Phi_{\text{dq}}}. \nonumber
\eea
In order to shield the phase accumulation from decoherence and thus
achieve a higher fidelity, we additionally implement a Hahn echo $\pi$
pulse $U_2 \otimes U_2$ in the middle of the free evolution period.
Phase disturbances due to any quasi-static detuning (e.g. hyperfine
interactions with \Nis nuclei or slow magnetic field variations) are dynamically decoupled by the echo, allowing
for a $T_2$-limited gate fidelity.
Taking into account the modest coherence time of NV1 ($T_{2 \text{dq}} = 150 \pm 17$~$\mu$s) and the
initial spin polarization (here 0.97 for each electron spin), the
theoretical upper bound for the gate fidelity is
$F_{\text{lim}} \approx 0.849$ which is in agreement with our measurement results.
In the previous work on generating entanglement between two NV centers~\cite{dolde_room_temperature_2013} the
fidelity was severely limited by pulse errors in the 16~local $\pi$
and $\pi/2$ pulses used in
the sequence, reducing it down to $F_{\text{std}} = 0.67 \pm 0.04$.
By replacing these 16 rectangular mw pulses by three numerically optimized
local gates we were able to improve the fidelity up to $F_{\text{opt}} = 0.824 \pm 0.015$ which reaches the limit set by decoherence and initialization fidelity.


Next we shall demonstrate entanglement storage on the nuclear spins
using the \SWAP{} gate introduced above.
To this end a control sequence was optimized to execute simultaneous
\SWAP{} gates on both NV centers yielding a storage efficiency of
$\text{Eff}_{\text{opt}}=0.92 \pm 0.07$ (compared to 
$\text{Eff}_{\text{std}}=0.39$ achieved with standard pulses in previous work
\cite{dolde_room_temperature_2013}).
The fidelity of the entangled state after storage and retrieval is $F_{\text{opt retrieved}} = 0.74 \pm 0.04$.
It is important to note that during the spin state storage the two
remote nuclear spins are entangled.
Using reconstructed electron spin density matrices
before, during and after the entanglement storage we may estimate
the fidelity of the entangled nuclear spin
state to be \mbox{$F_{\text{opt nuclear}} = 0.819$}.
The corresponding estimated density matrix of the entangled nuclear spins is shown in Figure~\ref{fig:4}c (see Supplementary Information).
In order to quantify the entanglement~$E(\rho)$ in our states,
we minimized the relative entropy
$S(\rho\|\sigma) := \trace\left(\rho(\log \rho -\log \sigma)\right)$
over all separable states~$\sigma$~\cite{Vedral_ent_measure},
yielding $E_{\text{electron}} = 0.37$
and $E_{\text{nuclear}} = 0.23$.
This demonstrates a significant improvement of the NV-NV electron
spin quantum correlation in comparison with standard control, yielding only
$E_{\text{std}} \approx 0.16$.

\begin{table}[t]
\centering
\begin{tabular}{rcccc}
\toprule
\textit{fidelity for various}&standard&optimized&calc. limit of&calc. limit with\\
\textit{control sequences}&control&control&current pair&optimum values$^*$\\
\midrule
\gate{not} gate & $0.94$ & 0.99 & $>0.99$ & $>0.999$\\
entangling sequence & $0.67 \pm 0.04$~\cite{dolde_room_temperature_2013} & $0.824 \pm 0.015$ & $0.849$ & $>0.993$\\
\SWAP{} gate & 0.87 & $0.97 \pm 0.01$ &0.97 & $ >0.999$\\
\bottomrule
\end{tabular}
\caption{\label{tab:1}
  \textbf{Fidelity comparison.}
The upper limits are due to imperfect initialization and dephasing during the sequence.
$^*$Optimum values refer to the current record values for initialization fidelity ($>0.99$), coherence lifetime ($T_2=4\,$ms) and spin state eigenbasis.
}
\end{table}


In conclusion, we have demonstrated that the implementation of optimal control is a prerequisite
for the realization of spin-based quantum information technology.
The implementation itself is perhaps more challenging than in many other
types of quantum systems due to the high level of crosstalk present in a multi-spin system.
Such crosstalk has been identified as a limiting feature
that needs to be overcome to make spin-based registers scalable.
The present study offers strong supporting evidence that this challenge can indeed be overcome by optimal control.
Especially for the nuclear spin storage (and thereby nuclear spin entanglement), crosstalk becomes a major issue.
Here, an entanglement \SWAP{} fidelity larger than $0.94\pm 0.03$ is demonstrated, enabling meaningful
entanglement storage and nuclear spin entanglement protocols.
In this setting, our work may thus be envisaged as a breakthrough, where optimal control
is an indispensable tool to achieve the combination of several
highly demanding tasks simultaneously: (a) high-end control of transitions in a crowed
spectrum with 36 energy levels, (b) suppression of crosstalk, (c) creation of entanglement
between distant nuclear spins with different quantization axes via control of electron-nuclear
interactions on several timescales, and (d) decoupling from unwanted interactions.
Our control methods, though tailored
for NV centers, can easily be transferred to other types of experimental
set-ups as well.
Thus they are anticipated to find wide application.
At the moment the performance is mainly limited by the coherence times of the electron spins.
However, this is a material property and long coherence times for artificially created NV centers have been demonstrated in
isotopically purified diamond  \cite{yamamoto_extending_2013}.
Recent advances in implantation techniques (i.e. low energy mask
implantations \cite{toyli_chip-scale_2010}) as well as coherence time extension by
growing an additional layer of diamond over the implanted NVs \cite{staudacher_enhancing_2012} will
pave the way for a high-yield chip size fabrication of NV
arrays. The methods developed in this work will play a crucial role in
making the control of such spin arrays feasible.
The control fidelity could be further improved by robust control
sequences which can automatically compensate for small magnetic field,
temperature and control power fluctuations.
Since the achieved control fidelity depends on the accuracy of the simulation
used in optimization, accurate measurement of the system parameters
(e.g. the hyperfine tensor) is of paramount importance.
In principle the pulses could also be improved using closed-loop
optimization where measurement data is immediately fed back to the
optimizer to improve the pulses without full knowledge of the system.

\section*{Methods}
The diamond sample is grown by microwave-assisted chemical vapor deposition (CVD).
The intrinsic nitrogen content of the grown crystal is below $1\,$ppb and the \Cis content is enriched to 99.9~\%.
\Nis ions were implanted with an energy of $1\,$MeV through nano-channels in a mica sheet.
A characterization of this method was published recently \cite{Pezzagna_mica_2011,dolde_room_temperature_2013}.

The two NV centers of this work are optically addressed by a home-built confocal microscope.
Microwave (mw) radiation was guided to the NV centers of interest using a lithographically fabricated coplanar waveguide structure on the diamond surface.
Microwave control was established with an home-built IQ mixer and an AWG (Tektronix AWG 5014C) to generate arbitrary microwave amplitudes, frequencies and phases.
With the microscope and mw devices optically detected magnetic resonance (ODMR) of single NV electron spins is performed.
To this end the laser is used to initialize the electron spin into its $\ket{0}$ state by laser excitation and subsequent decay.
Next, the spin is manipulated by mw fields.
Finally, the fluorescence response to a next laser pulse reports on the spin state (i.e. low level for $\ket{\pm}$ and high level for $\ket{0}$).

The $S=1$ electron spin of the NV center experiences a strong crystal
field of about 100~mT along the center's symmetry axis,
splitting apart the $\ket{\pm}$ levels from~$\ket{0}$.
As the symmetry axis has four possible orientations in a diamond crystal lattice, NV centers might differ in crystal field direction as for the present NV pair.
A small magnetic field is used to lift the remaining degeneracy of $\ket{\pm}$ to guarantee individual addressing of spin transitions.
Here, using magnetic field coils a magnetic field of $3.41$~mT with an angle of about $24^{\circ}$ to the NV1 axis and $125^{\circ}$ to the NV2 axis was applied.
In order to have no effect from the different charge states of the NV center charge state pre-selection was implemented \cite{waldherr_violation_2011}

The optimization of the control sequences was performed
using a customized version of the \DYNAMO{}~\cite{dynamo} optimal control
framework, which utilizes the GRAPE~\cite{GRAPE} algorithm to
compute the gradient of the error function, and a standard
numerical optimization algorithm (such as BFGS) to minimize it.
Crosstalk was handled using a modified rotating wave approximation (RWA),
taking sufficiently slowly rotating Hamiltonian components into account.
More details can be found in the Supplementary Information.

\section*{Acknowledgments}
The authors would like to acknowledge financial support by the EU via SQUTEC and Diamant, as well as the DFG via the SFB/TR21, the research groups 1493 ``Diamond quantum materials'' and 1482 as well as the Volkswagen Foundation.
Moreover, the work was supported in part by the DFG through SFB631, by the EU integrated projects QESSENCE and SIQS as well as by the Bavarian Excellence
Network (ENB) through QCCC.  Federica Ferraris assisted in the preparation of Fig~1a.

\bibliography{nv}

\newpage
\newpage
\newpage

\newpage
\begin{figure}
  \includegraphics[width=0.5\textwidth]{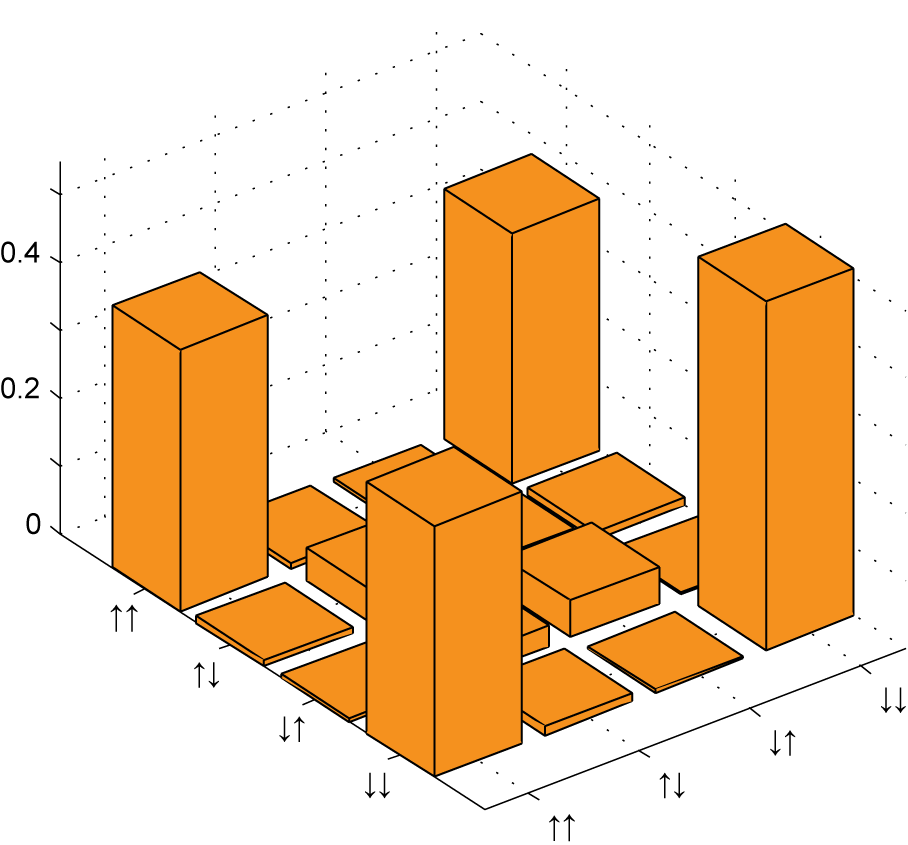}
  \caption{\label{fig:4}
    \textbf{Nuclear spin entanglement.}
    Using optimal control we have swapped the entangled electron spin state onto the nuclear spins (F=$0.819$).
		The orange density matrix represents the entangled Bell state $\ket{\Phi} = \ket{\uparrow \uparrow} -i\ket{\downarrow \downarrow}$ of the two remote nuclear spins.
  }
\end{figure}

\appendix
\section{Hamiltonian of the NV system}

\newcommand{\carrier}{\tilde{\omega}}
\newcommand{\larmor}{\omega}
\newcommand{\rabi}{\Omega}

A single $^{15}$NV$^-$ center in a static magnetic field
$\vec{B}_0 = B_0 \vec{u}_0$
has the Hamiltonian
\begin{align}
\label{eq:Hsingle}
\notag
H/\hbar
&= 2 \pi \Delta S_z^2
-\gamma_e \vec{B}_0 \cdot \vec{S}
-\gamma_N \vec{B}_0 \cdot \vec{I}
+2 \pi \vec{S} \cdot A \cdot \vec{I}\\
&= 2 \pi \Delta S_z^2
+\larmor_e \vec{u}_0 \cdot \vec{S}
+\larmor_N \vec{u}_0 \cdot \vec{I}
+2 \pi \sum_k A_{kk} S_k I_k,
\end{align}
where $\vec{S}$ and $\vec{I}$ are the dimensionless spin operators for
the electron pair and the \Nis nucleus, respectively, quantized
along the NV symmetry axis.
Lattice strain has been neglected.
$\Delta \approx 2.87$~GHz is the zero-field splitting.
The anisotropic (but axially symmetric) hyperfine coupling coefficients are
$A_{xx} = A_{yy} \approx 3.65$~MHz and
$A_{zz} \approx 3.03$~MHz~\cite{felton2009}.
The Larmor frequencies are defined as~$\larmor_i := -\gamma_i B_0$,
where $\gamma_i$ is the gyromagnetic ratio of the spin (electron or nuclear).
In a typical experiment $\larmor_e \approx 100$~MHz.

The system can be controlled using oscillating magnetic fields of the form
\be
\label{eq:Bcontrol}
\vec{B}_k(t) = B_k(t) \cos(\carrier_k t +\phi_k(t)) \vec{u}_k,
\ee
where $\carrier_k$ are the carrier frequencies (in the microwave region).
The amplitudes $B_{k}$ and the phases $\phi_k$ can be changed in time
to steer the system.
The unit vectors $\vec{u}_k$ representing the polarization of the
control signal are determined by the antenna setup.
In our case $\vec{u}_k \parallel [001]$.
The control fields add additional Zeeman terms for both the electron and
the nuclear spins:
\be
\label{eq:HC}
H_k(t)/\hbar = -\vec{B}_k(t) \cdot (\gamma_e \vec{S} +\gamma_N \vec{I})
= \underbrace{\frac{-\gamma_e B_k(t)}{\sqrt{2}} |\vec{u}_k^\perp|}_{\rabi_k(t) :=}
  \underbrace{\sqrt{2} \frac{\vec{u}_k}{|\vec{u}_k^\perp|} \cdot \left(\vec{S}
  +\frac{\gamma_N}{\gamma_e} \vec{I}\right)}_{C_k :=} \cos(\carrier_k t +\phi_k)
= \rabi_k(t) C_k \cos(\carrier_k t +\phi_k),
\ee
where $\rabi_{k}(t)$ is the \emph{driving Rabi frequency} and
$C_k$~the corresponding \emph{control operator}.
The reason for the strange normalization is that when $\vec{B_0}$ is
aligned with the NV axis, only
the perpendicular component of the control field drives a population transfer.

The system of two coupled NV centers is then described by the Hamiltonian
\be
\label{eq:Hfull}
H = H_A + H_B + H_{\text{int}},
\ee
where $H_A$ and $H_B$ are the Hamiltonians of the individual NV
centers, NV-A and NV-B, respectively, and $H_{\text{int}}$ describes the
dipolar interaction between them:
\be
\label{eq:Hdip}
H_{\text{int}}/\hbar
= \frac{\mu_0}{4\pi}\frac{\hbar \gamma_e^2}{r_{AB}^3}
\left(\vec{S}_A \cdot \vec{S}_B - 3(\vec{S}_A \cdot \hat{r}_{AB})(\vec{S}_B \cdot \hat{r}_{AB})\right).'
\ee
The two NV centers are separated by a distance of
$r_{AB} = 25 \pm 2$~nm, and the strength of the dipole-dipole interaction between them is found to be
$\nu_{\text{dip}} = 4.93 \pm 0.05$~kHz.
Because of the strong, local zero field splitting and Larmor terms, the
effect of all the $H_{\text{int}}$ terms but the ${S_z}_A {S_z}_B$ one are strongly
suppressed and may be neglected. Thus we obtain
\be
H_{\text{int}}/\hbar \approx 2 \pi \nu_{\text{dip}} {S_z}_A {S_z}_B.
\ee
The dipolar interactions between $\vec{S}_{A/B}$ and $\vec{I}_{B/A}$, and
between the two nuclear spins are weaker by factors of
$\frac{\gamma_e}{\gamma_N} \approx 6500$ and $(\frac{\gamma_e}{\gamma_N})^2$, respectively,
and can be safely ignored.

\begin{figure}
  \includegraphics[width=0.5\columnwidth]{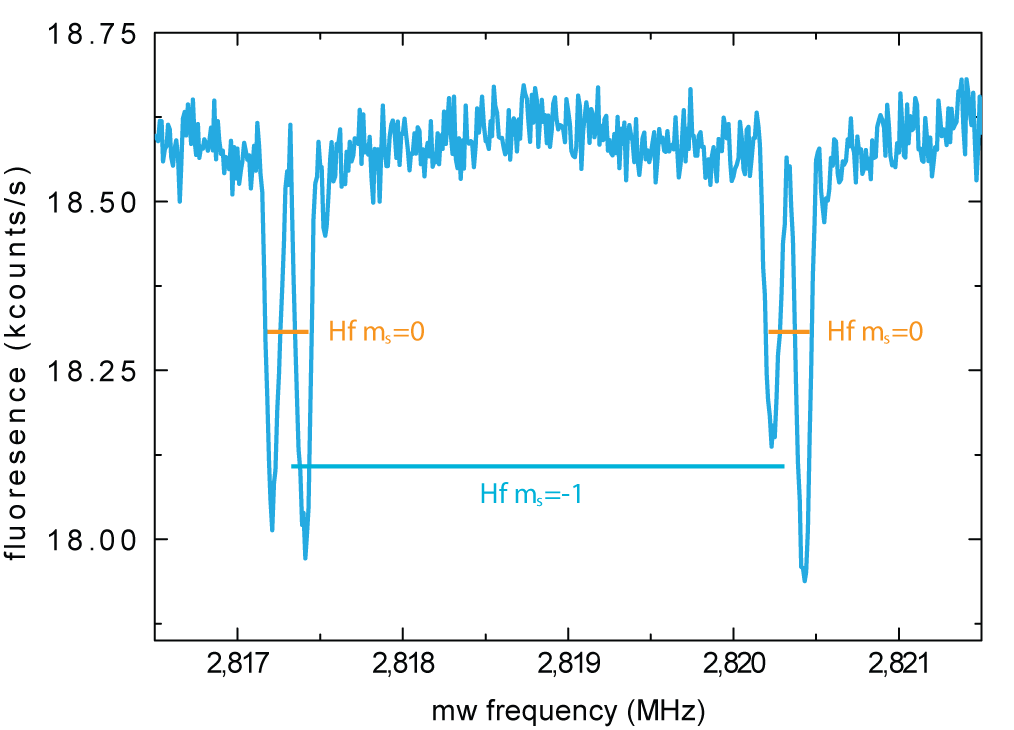}
  \caption{\label{fig:odmr}
    \textbf{High resolution ODMR spectrum.}
     The spectrum was recorded for the $m_{s}=0 \to m_{s}=-1$ transition of~NV2.
  }
\end{figure}

In the experiment, the two NV centers have different axis
orientations, $[111]$ and $[\bar{1}1\bar{1}]$, which makes them
individually addressable even in a uniform magnetic field.
The static magnetic field $\vec{B}_0$ makes the angle
$\theta_{A} \approx 0.133 \pi$
with~$\hat{z}_A$ and the angle
$\theta_{B} \approx 0.695 \pi$
with~$\hat{z}_B$. 
Such alignment leads to considerable hyperfine splitting in the
$m_S=0$ level (see Fig.~\ref{fig:odmr}) due to a small admixture of
levels $m_S=\pm1$ which lead to small magnetic moment roughly
perpendicular to the NV axis.
As the hyperfine field at the nitrogen nucleus for the $m_S=0$ level is almost perpendicular to the ones for levels $m_S=\pm1$ different nuclear spin quantization axes arise.
The latter can be utilized for coherent nuclear spin control in the in $m_s = 0$ subspace via the electron spin, e.g.
to perform a (partial) swap operation between the electron spin and the nuclear spin.

\section{Calibration}

Once the Hamiltonian parameters are known (by fitting them to the
measured hyperfine ODMR peaks such as the ones in Fig.~\ref{fig:odmr})
we determine the (in general nonlinear) dependence between the
amplifier setting~$a$ and the corresponding driving Rabi
frequency~$\rabi_k$ for each carrier frequency~$\carrier_k$ separately.
This is done by finding, for a set of values of~$a$, the $\rabi_k$:s that yield the best
match between simulated and measured single driving data, and doing
e.g. monotonous cubic spline interpolation between the points.

\section{Rotating frame approximation}

We use two independent methods to simulate our system. Both yield high
fidelity pulses. One approach is to apply perturbation theory
first to remove non-secular terms in the free evolution Hamiltonian~\cite{childress_coherent_2006}.
When moving into a rotating frame the control Hamiltonian still has time-dependent terms,
which can be made time-independent by using Floquet theory~\cite{floquet_theory_2010}.
The second approach directly employs the rotating frame
approximation and then drops any terms with a small
amplitude-to-rotation-frequency ratio. Here we will describe the second method in
detail.

A rotating frame is an interaction picture defined by a
time-independent, typically local, Hamiltonian~$H_0$.
Given a system with the Hamiltonian~$H$, we have in the Schrödinger picture
$i \hbar \partial_t \ket{\psi} = H \ket{\psi}$.
We then define the interaction picture ket
\be
\ket{\psi'} := \underbrace{e^{i H_0 t/\hbar}}_{U_0(t)} \ket{\psi}.
\ee
At $t=0$, the rotating frame coincides with the lab frame.
The corresponding transformation for operators is
$A' := U_0(t) A U_0^\dagger(t)$.

Assume $H_0$ has the spectral decomposition
$H_0/\hbar = \sum_k \omega_k P_k$,
where $\omega_k$ are unique and arranged in increasing order,
and the orthogonal eigenspace projectors~$P_k$ sum to identity.
Now
\be
\label{eq:rw_projectors}
A'
= e^{i H_0 t/\hbar} \left(\sum_a P_a\right) A \left(\sum_b P_b\right) e^{-i H_0 t/\hbar}
= \sum_{ab} e^{i (\omega_a-\omega_b) t} P_a A P_b.
\ee

Assume that the system Hamiltonian is of the form
\be
H = H_0 +\sum_k \rabi_k(t) C_k \cos(\carrier_k t +\phi_k),
\ee
where the carrier frequencies $\carrier_k \ge 0$ by convention.
The rotating frame Hamiltonian is given by
\begin{align}
\label{eq:RWA}
\notag
H'
&= \sum_{kab} \rabi_k(t) P_a C_k P_b e^{i (\omega_a-\omega_b) t} \cos(\carrier_k t +\phi_k)\\
&=
\sum_{ka} \rabi_k(t) P_a C_k P_a \cos(\carrier_k t +\phi_k)
+\frac{1}{2} \sum_{k,a<b} \rabi_k(t) \left(P_a C_k P_b
\left(e^{i (\omega^\text{slow}_{kab} t +\phi_k)} +e^{-i (\omega^\text{fast}_{kab} t +\phi_k)}\right)
+\text{h.c.} \right),
\end{align}
where $\delta_{ab} := \omega_a-\omega_b$, and we have further defined
$\omega^\text{slow}_{kab} := \carrier_k +\delta_{ab}$ and
$\omega^\text{fast}_{kab} := \carrier_k -\delta_{ab}$.

We use Eq.~\eqref{eq:RWA} to approximate the
rotating frame Hamiltonian $H'$ using the static term and a small number
of slowly rotating terms. For each carrier frequency, all the terms
which rotate at the same frequency $\omega$
(collected in the ordered pair index set $Q(\omega)$)
are added together and retained if
\be
s \: \rabi_{\text{max}} \left\|\sum_{(a,b) \in Q(\omega)} P_a C P_b\right\|_{\text{F}} > |\omega|,
\ee
where $s = 300$ is a cutoff parameter.
The maximum control amplitude $\rabi_{\text{max}}$ is chosen such that no
fast mode is kept.

We apply the control microwaves at four distinct carrier frequencies, each in the center of
the observed hyperfine peaks of a single-NV $\ket{0} \to \ket{+}$ or $\ket{0} \to \ket{-}$
transition.
A convenient rotating frame is thus obtained by choosing
$H_0$ to consist of the electron Zeeman and zero-field
splitting terms, which makes the highest-magnitude control terms static.
However, because of the relatively high spectral transition density in the
NV-NV system we will have some crosstalk, manifesting
itself as non-negligible slowly rotating terms in the rotating frame Hamiltonian~$H'(t)$
which need to be taken into account.

Since an off-axial $\vec{B}_0$ field makes $H_0$ slightly non-diagonal,
the $U_0(t)$ transformation does not keep our observable
$O = a\ket{0}\bra{0}_A +b\ket{0}\bra{0}_B +c \I$ perfectly invariant
in time. This introduces a small additional error to the measurement.

\section{Numerical pulse optimization}

In order to implement a high-fidelity quantum gate~$G$, i.e.~a specific unitary
propagator of the system, we resort to optimal control techniques.
The procedure involves defining an equation of motion for the system
(in our case the Schrödinger equation in a rotating frame under the
Hamiltonian in Eq.~\eqref{eq:Hfull}),
a set of control fields (the driving Rabi frequencies~$\rabi_k(t)$ and
the phases~$\phi_k(t)$ in Eq.~\eqref{eq:HC}), and a cost functional to
be numerically minimized.
For reasons of computational efficiency and ease of implementation
the control fields are taken to be piecewise constant in time.
The cost functional is simply the error function
\be
\label{eq:err}
E[\rabi_k(t), \phi_k(t), T] = 1 -\frac{1}{D} \left|\trace(G^\dagger
U(T))\right| \quad \in [0, 1],
\ee
where $U(T)$ is the propagator obtained by integrating the Schrödinger
equation of the system from $0$ to~$T$ under the control sequence,
and $D$ the total dimension of the system. This choice of error
function automatically absorbs unphysical global phases.

In some cases we are only interested in what happens to a specific
subsystem, i.e. we wish to obtain a propagator of the form
$G \otimes W$ where $G$ is the gate to be implemented and $W$ is an
arbitrary unitary.
The fact that we do not care what happens to the other subsystem(s) as long as the
total propagator remains factorizable can make the optimization task much simpler.
In this case the appropriate error function is
\be
E_{\text{partial}}[\rabi_k(t), \phi_k(t), T] = 1 -\frac{1}{D}
\left\|\trace_1((G \otimes \I)^\dagger U(T))\right\|_{\text{tr}} \quad
\in [0, 1],
\ee
where the trace norm $\|A\|_{\text{tr}} = \sum_k \sigma_k$ is given by
the sum of the singular values of~$A$.
It is easy to see how this reduces to Eq.~\eqref{eq:err} when the
second subsystem is trivial.

Due to the rapid oscillation of the control
Hamiltonian~\eqref{eq:HC} it is much faster to perform the
integration in a suitable rotating frame, discarding all the non-static
terms in the rotating frame Hamiltonian and thus making it time
independent. This way we may utilize the GRAPE~\cite{GRAPE} algorithm to efficiently
compute the gradient of the error function, and a standard
optimization algorithm (such as BFGS) to minimize it, using a customized
version of the \DYNAMO{}~\cite{dynamo} optimization framework.
However, this is an approximation which does not
take into account crosstalk, which in our case can be significant.
In order to push the gate fidelity higher it needs to be accounted for.
Hence we only use the fast, rough method in the initial phase of the
optimization. Once the gate error is low enough we switch over to a
more accurate time-dependent rotating frame Hamiltonian which includes
slowly rotating terms representing the most significant crosstalk components.

The fidelities of the control sequences obtained in this way are
ultimately limited by (1) the accuracy of the simulation, (2) the
approximations used, and (3) decoherence.
The specific decoherence mechanisms can also be included in the optimization,
but in our scenario (generation of full quantum gates) we did not deem it worthwhile.

\section{Entangling sequence and \SWAP{}}

The electron spins of the two NV centers can be entangled using a non-local
phase gate generated by~$H_{\text{int}}$.
The sequence used in the experiment is
\bea
\ket{00} & \xrightarrow{U_1 \otimes U_1} & \frac{1}{2}\left(\ket{+}+\ket{-}\right) \otimes \left(\ket{+}+\ket{-}\right) \nonumber \\
& \xrightarrow{e^{-i H_{\text{int}} \tau/\hbar}} & \frac{1}{2} \left((\ket{++}+\ket{--}) +e^{i \phi}(\ket{+-}+\ket{-+})\right)\\
& \xrightarrow{U_3 \otimes U_3} &
\frac{1}{\sqrt{2}}\left(\ket{++}+i\ket{--}\right) =: \ket{\Phi_{\text{dq}}}. \nonumber
\eea
where the free evolution time $\tau = \frac{1}{8 \nu_{\text{dip}}}
\approx 25.4\,\mu$s will yield
the relative phase $\phi = \pi/2$, and the local gates are defined as
\be
U_1 =
\frac{1}{\sqrt{2}}
\begin{pmatrix}
i & 1 \\
& &  \sqrt{2}\\
-i & 1
\end{pmatrix},
\quad
U_2 =
\begin{pmatrix}
& & 1\\
& 1\\
1
\end{pmatrix},
\quad
U_3 =
\frac{1}{\sqrt{2}}
\begin{pmatrix}
-1 & & 1\\
& \sqrt{2}\\
1 & & 1
\end{pmatrix}.
\ee
The $U_2 \otimes U_2$ gate is used to implement a Hahn echo in the
middle of the free evolution period to cancel unwanted phases accumulated
due to quasi-static magnetic field noise, and the hyperfine interactions.

Nuclear spins couple to magnetic fields much more weakly than the
electrons, and consequently have much longer coherence times.
Therefore they can be used as a low-decoherence storage space for
quantum states prepared using the electron spins.
The storage and retrieval happens
by swapping the nuclear spin state with that of the electron.
For \Nis the corresponding Hilbert spaces have different
dimensions so a full \SWAP{} gate is not possible.
Instead we will use a partial \SWAP{} gate which exchanges the states
$\ket{+\uparrow}$ and $\ket{-\downarrow}$, and keeps the other four
states invariant without introducing any unwanted phases:
\be
\textsc{swap} =
\left(
\begin{array}{cc|cc|cc}
0 & & & & & 1\\
  & 1 & & & & \\ \hline
  & & 1 & & & \\
  & & & 1 & & \\ \hline
  & & & & 1 & \\
1 & & & & & 0
\end{array}
\right)
\ee
This is enough to store the Bell-type entangled state~$\ket{\Phi_{\text{dq}}}$ obtained
in the previous part of the experiment.
The gate is also self-inverse, $\textsc{swap}^2 = \I$, which means we can
use the same control sequence for both storage and retrieval.


\section{Nuclear spin entanglement}

\begin{figure}
  \includegraphics[width=\columnwidth]{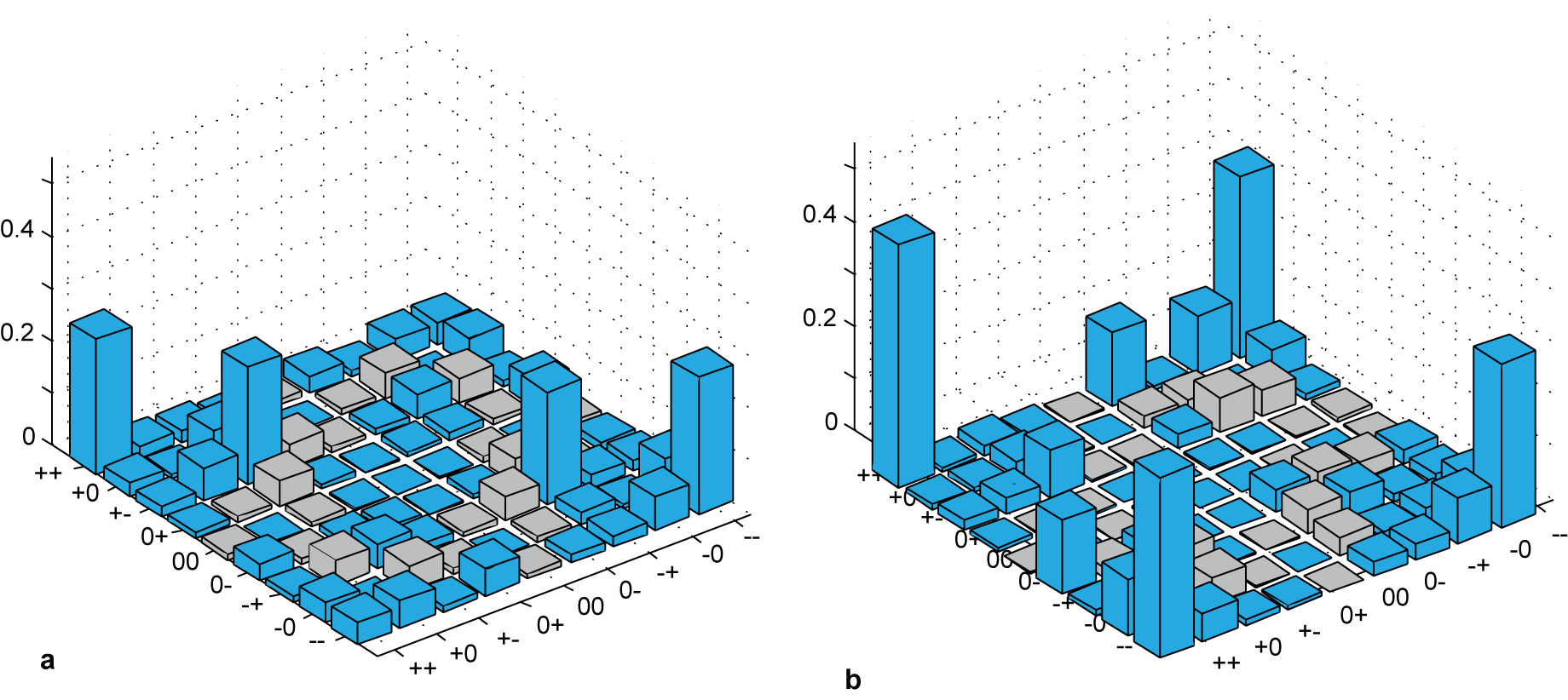}
  \caption{\label{fig:tomo}
    \textbf{Electron state tomography.}
    Reconstructed electron spin density matrices.  (a) $\hat{\rho}_B$, after
    entanglement storage. (b) $\hat{\rho}_C$, after entanglement retrieval.
  }
\end{figure}

Although we can only directly measure the electron spins, it is
possible to estimate the nuclear spin state via its hyperfine coupling
to the electron spin, which generates the \SWAP{} gate we use for
entanglement storage and retrieval.
Let us denote the electron state tomographies after
the entangling sequence, entanglement storage, and retrieval by $\hat{\rho}_A$, $\hat{\rho}_B$
and $\hat{\rho}_C$, respectively.
The tomographies are presented in Fig.~\ref{fig:tomo}.

According to simulation the full-system state $\rho_A$ is very close
to being factorisable, with
$F(\rho_A, \trace_e(\rho_A) \otimes \trace_{N1}(\rho_A) \otimes \trace_{N2}(\rho_A)) = 0.984$.
To obtain an estimate for the nuclear spin state after entanglement
storage, we will set
$\sigma_A := \hat{\rho}_A \otimes \rho_{N1} \otimes \rho_{N2}$, and
minimize the error function
\be
E = \left\||S \sigma_A S^\dagger|-\hat{\rho}_B\right\|^2
+\left\| |S^2 \sigma_A S^{\dagger 2}|-\hat{\rho}_C\right\|^2
\ee
over all possible nuclear states $\rho_{N1}$ and $\rho_{N2}$.
The $S$~gate is the (imperfect) \SWAP{} gate obtained by simulating the
\SWAP{} control sequence.
The element-wise absolute value $|\cdot|$ is used because the upper limits in our
tomographies contain no phase information.

\section{Entanglement measure}

To estimate the amount of entanglement in a given state~$\rho$ we use the
entanglement measure introduced in~\cite{Vedral_ent_measure}, defined as
\be
E(\rho) :=
\min_{\sigma \in \mathcal{D}} S(\rho \| \sigma),
\ee
where $\mathcal{D} \subset \mathcal{H}$ is the set of all separable states.
The relative entropy $S(\rho \| \sigma)$ is given by
\be
S(\rho \| \sigma) := \trace\left(\rho (\log \rho -\log \sigma)\right),
\ee
and is taken to be infinite iff $\Supp \rho \cap \Ker \sigma \neq \{0\}$.

Essentially $E(\rho)$ measures the relative-entropy
``distance''\footnote{Strictly speaking, relative entropy is not a distance measure because it is not symmetric.}
of $\rho$
from the set of separable states. In practice it is estimated by
starting with
$\sigma_0 := \diag(\diag(\rho))$
and then generating a sequence of random separable states~$\zeta_k$,
in each iteration setting
$\sigma_{k} := (1-s_k) \sigma_{k-1} +s_k \zeta_k$,
where $s_k \in [0,1]$ minimizes~$S(\rho \| \sigma_k)$.
This process yields a strict upper limit for~$E(\rho)$.

For the ideal state
$\ket{\Phi_{\text{dq}}} = \frac{1}{\sqrt{2}}\left(\ket{++}+i\ket{--}\right)$
we may obtain analytically $E(\ket{\Phi_{\text{dq}}}) = \log 2$.

\end{document}